\documentclass[apjl]{emulateapj}
\usepackage{graphicx}
\usepackage{amssymb}
\usepackage{epstopdf}
\usepackage{color}

\shorttitle{Irregular cold fronts in NGC~7618/UGC~12491}
\shortauthors{Roediger et al.~2012}

\begin{document}

\newcommand{\CF}{_{\mathrm{CF}}}
\newcommand{\Hot}{_{\mathrm{hot}}}
\newcommand{\Cold}{_{\mathrm{cold}}}
\newcommand{\Max}{_{\mathrm{max}}}
\newcommand{\ICM}{_{\mathrm{ICM}}}
\newcommand{\KeV}{\,\textrm{keV}}
\newcommand{\Kpc}{\,\textrm{kpc}}
\newcommand{\Kms}{\,\textrm{km}\,\textrm{s}^{-1}}
\newcommand{\K}{\,\mathrm{K}}
\newcommand{\ccm}{\,\mathrm{cm}^{-3}}
\newcommand{\gccm}{\,\mathrm{g}\,\mathrm{cm}^{-3}}
\newcommand{\cmss}{\,\mathrm{cm}\,\mathrm{s}^{-2}}
\newcommand{\Myr}{\,\mathrm{Myr}}
\newcommand{\Gyr}{\,\mathrm{Gyr}}

\newcommand{\etal}{et al.}

\definecolor{rred}{rgb}{1,0,0}
\definecolor{bblue}{rgb}{0,0,1}

\title{Irregular sloshing cold fronts in the nearby merging groups NGC~7618 and UGC~12491:  evidence for Kelvin-Helmholtz instabilities}
\author{E.~Roediger\altaffilmark{1,2}, R.~P.~Kraft\altaffilmark{2}, M.~E.~Machacek\altaffilmark{2}, W.~R.~Forman\altaffilmark{2}, P.~E.~J.~Nulsen\altaffilmark{2}, C.~Jones\altaffilmark{2},  \& S.~S.~Murray\altaffilmark{3,2}}
\affil{
\altaffilmark{1}Jacobs University Bremen, Campus Ring 1, 28759 Bremen, Germany\newline
\altaffilmark{2}Harvard/Smithsonian Center for Astrophysics, 60 Garden Street, Cambridge, MA 02138, USA\newline
\altaffilmark{3}Department of Physics and Astronomy, The Johns Hopkins University, 3400 N. Charles St., Baltimore, MD 21218, USA
}
%\affil{}
\email{e.roediger@jacobs-university.de}

\begin{abstract}
We present results from two $\sim$30 ks Chandra observations of the hot atmospheres of the merging galaxy groups centered around NGC~7618 and UGC~12491.  Our  images show the presence of arc-like sloshing cold fronts wrapped around each group center and $\sim 100\Kpc$ long spiral tails in both groups.  
Most interestingly, the  cold fronts are highly distorted in both groups, exhibiting `wings' along the fronts. These features resemble the structures predicted from non-viscous hydrodynamic simulations of gas sloshing, where Kelvin-Helmholtz instabilities (KHIs) distort the cold fronts. This is in contrast to the structure seen in many other sloshing and merger cold fronts, which are smooth and featureless at the current observational resolution.  
Both magnetic fields and viscosity have been invoked to explain the absence of KHIs in these smooth cold fronts, but the NGC~7618/UGC~12491 pair are two in a growing number of both sloshing and merger cold fronts that appear distorted. 
Magnetic fields and/or viscosity may be able to suppress the growth of KHIs at the cold fronts in some clusters and groups, but clearly not in all.
We propose that the presence or absence of KHI-distortions in cold fronts can be used as a measure of the effective viscosity and/or magnetic field strengths in the ICM.  \end{abstract}

\maketitle

%****************
\section{Introduction} \label{sec:intro}
%****************
%
Minor mergers are a common phenomenon in the growth of galaxy groups and clusters and leave characteristic observable features in the intra-cluster medium (ICM). The passage of a less massive subcluster through the main cluster offsets the central gas peak from the central potential well.  Subsequently, the offset ICM falls back towards the potential minimum and thus oscillates -- or sloshes -- inside the cluster center (\citealt{Markevitch2001,Ascasibar2006}). The transfer of angular momentum in off-center mergers gives the sloshing gas a spiral-like appearance. The sloshing forms cold fronts (CFs) in the gas,  i.e.~arc-shaped discontinuities in density, temperature, metallicity and thus X-ray brightness wrapped around cluster cores. Similar features can occur in equal-mass mergers if the merger partners pass each other at a sufficiently large distance such that their cores are only disturbed but not destroyed.  Sloshing  CFs are commonly seen in clusters (\citealt{Markevitch2003,Ghizzardi2010}) and  have been observed in detail in several clusters and groups (see also review by \citealt{Markevitch2007}), 
e.g.~Abell 1795 (\citealt{Markevitch2001,Bourdin2008}), 
Abell 2142 (\citealt{Markevitch2000,Owers2009hifid}), 
Abell 496 (\citealt{Dupke2007}), 
NGC 5098 (\citealt{Randall2009ngc5098}),
NGC 5044 (\citealt{Gastaldello2009}),
NGC 6868 (\citealt{Machacek2010}),
NGC 5846 (\citealt{Machacek2011}), 
the Perseus cluster (\citealt{Churazov2003,Sanders2005perseus}), 
and the Virgo cluster (\citealt{Simionescu2010}).
\citet{Roediger2011,Roediger2012a496} performed hydrodynamic merger simulations specifically tailored to the Virgo cluster and Abell 496. By varying the merger geometry and mass ratio, the observed positions of the CFs and contrasts across them can be reproduced quantitatively and thus the recent merger history be constrained.

Sloshing CFs have generally been considered as smooth arcs (e.g.~in Virgo, Abell 2142). This is in contrast to predictions from high-resolution purely hydrodynamic simulations, where the CFs are distorted by Kelvin-Helmholtz instabilities (KHIs)  due to shear flows along the front (\citealt{ZuHone2010,Roediger2011,Roediger2012a496}). Both viscosity and magnetic fields have been invoked to explain the absence of KHIs in observed CFs (\citealt{Keshet2010,ZuHone2011}). However, some CFs do have significant substructure.   For example, the multiple sloshing CFs in A496 clearly display a remarkable boxy morphology  which had been unexplained until recently. Our high-resolution non-viscous simulations (\citealt{Roediger2012a496}) reproduce these boxy features and demonstrate that they arise due to KHIs at the CFs plus their interplay with projection. Also the sloshing CF in RX J1720 (\citealt{Markevitch2007}) is somewhat boxy-shaped, and the CFs in NGC~7618 and  UGC~12491 presented in this paper are clearly distorted. 

NGC~7618 and  UGC~12491 are the central galaxies of two nearby, roughly equal mass galaxy groups ($z$=0.017309, $d_L$=74.1 Mpc, 1$'$ $= 21\Kpc$).  An ASCA/GIS image of the pair (\citealt{Kraft2006}) shows extended X-ray emission centered on each of the dominant early-type galaxies. A short Chandra observation of NGC~7618 (\citealt{Kraft2006}) found a CF $\sim 20\Kpc$ north of the nucleus and a spiral-like tail.  
We obtained new 30 ks Chandra/ACIS-S images of the central regions of both groups 
that confirm the CF plus spiral tail structure in NGC~7618 and find a  similar morphology in UGC~12491. These CFs, the clear spiral-shaped gas asymmetries or tails in both groups and their nearly identical recessional velocities \citep{Huchra1999} suggest a group/group merger taking place almost entirely in the plane of the sky. This favorable viewing geometry and the proximity of the groups  make this an ideal target to study the gas dynamics of the merger process. 

In this letter, we discuss the distorted morphologies of the CFs present in both galaxies and the KHI as the likely origin of the distortions. 
A detailed analysis of the dynamic and thermodynamic state of the gas in this pair will be presented in a separate publication 
(M. Machacek \etal, in preparation). 
This paper is organized as follows.  Section 2 contains a short description of our data.
In section 3, we discuss the merger history of this pair of groups. Finally, in section 4, we present the distorted morphology of the CFs and discuss the implications of our results.

%**********
\section{Data}
%**********

NGC~7618 and UGC~12491 were observed for 34.5 and 33.2 ks with Chandra/ACIS-S (OBSIDs 7895 and 7896), respectively, as part of the HRC Guaranteed Time Observation program (PI:  S. S. Murray). The data were filtered for periods of high background.  Intervals where the rate in the 10-12 keV band was more than 3$\sigma$ above the mean rate were removed. After filtering, 32.15 and 31.1 ks of good time remained for NGC~7618 and UGC~12491, respectively. Bad pixels, hot columns,  columns along node boundaries and point sources were also removed. All data were processed with CIAO version 4.3.
We used blank sky background sets appropriate for the observation date and 
instrument configuration renormalized to agree with the observations 
in the $10-12$\,keV energy band, where particle background is expected 
to dominate. 
For more detail on the reduction and analysis
procedures, see Machacek et al., in preparation.

%************
\section{The merger between the two groups} \label{sec:results}
%************

%FFFFFFFFFF
\begin{figure}
\begin{center}
\includegraphics[trim=0 0 0 40,clip,angle=0,width=0.48\textwidth]{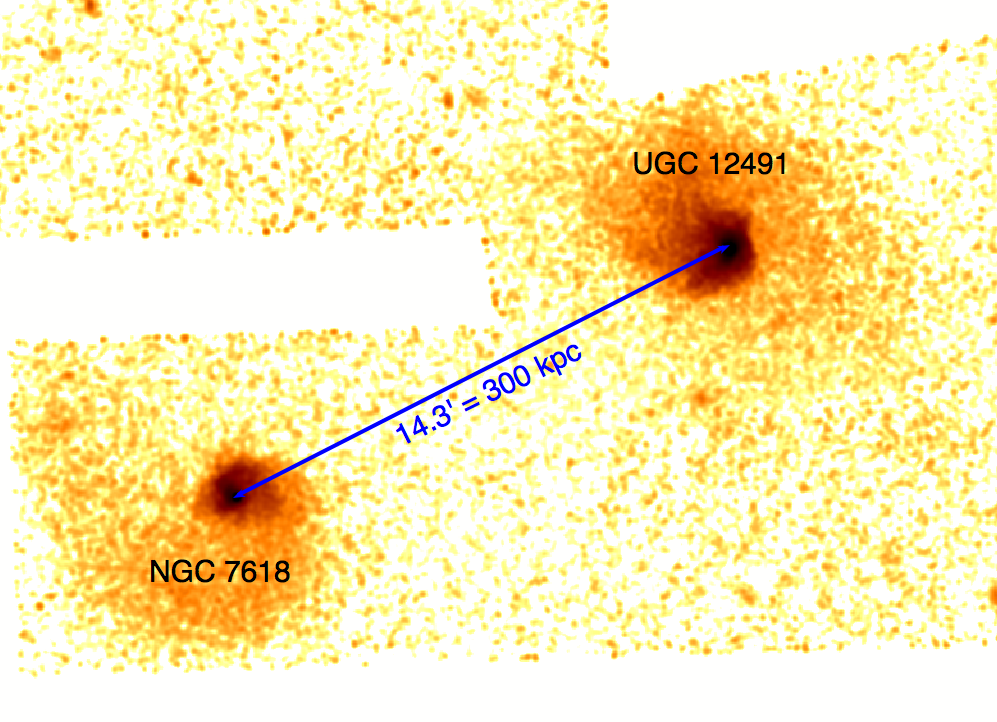}
\caption{Two co-added, background-subtracted and exposure corrected 30 ks  Chandra/ACIS-S images of the NGC~7618 and UGC~12491 groups in the 0.5-2.0 keV band, smoothed with a 4 arcsec Gaussian kernel.}
\label{fig:largescale}
\end{center}
\end{figure}
%FFFFFFFFFF

A mosaic of the NGC~7618 and UGC~12491 field is shown in Figure~\ref{fig:largescale}.  Both groups show characteristic gas sloshing features (see \citealt{Roediger2012a496} for a detailed list), i.e.~the arc-shaped CFs around the cores in a spiral-like fashion, and the spiral-shaped brightness excesses (or spiral tails) extending to several CF radii. NGC~7618 shows a clear arc-shaped surface brightness discontinuity 20 kpc to the north of the nucleus, and a spiral tail that curves from the south through the east and north.  UGC~12491 exhibits a very similar structure but at a rotated orientation -- a surface brightness discontinuity to the south-west of the nucleus and a spiral tail curving from  east over north then west. Temperature measurements using the mean energy  in the Fe L peak ($0.7 \lesssim E \lesssim 1.2$\,keV) as a proxy of temperature (\citealt{David2009}) confirm the brighter side of the surface brightness discontinuities to be the cooler one and thus the discontinuities to be CFs. 
The morphological comparison of the orientation of the CFs and the spiral tails to sloshing simulations \citep{Roediger2011,Roediger2012a496} suggests that both galaxy groups passed each other such that  
UGC~12491 came from the east-south-east and passed south-west of NGC~7618.

We estimate the time since the closest passage between the groups in two ways.  First, 
 the almost identical redshifts of both group centers indicate a merger in the plane of the sky. The average group gas temperatures are $\sim 1.2$ keV, corresponding to a sound speed of $560 \Kms$.
Around pericenter passage in cluster mergers, the relative velocity between both clusters varies between Mach 1 and 2 (e.g.~\citealt{ZuHone2010,Roediger2011}), merger partners on orbits with close encounters reach up to Mach 3 at pericenter passage. In the case of the NGC 7618/UGC 12491 merger, both groups are of similar mass, but their cores have not been destroyed but only disturbed, which argues against a close encounter. Hence, assuming an average relative velocity  between Mach 1.5 ($840\Kms$)  and Mach 2 ($1120\Kms$)   since their closest approach, the projected distance of 300 kpc translates into an age of around 0.3 Gyr. 
Second, as the outwards motion of sloshing CFs is mainly governed by the host potential, the distance of the CFs from the center depends mainly on the time since the closest encounter. Although the velocity of the CFs can in principle  differ from cluster to cluster, \citet{Roediger2011,Roediger2012a496} found  similar CF velocities for Virgo and Abell 496. Only in more massive clusters, like A2029, do the CFs progress out faster (\citealt{Roediger2012fastslosh}). In all simulated systems, the CFs move with approximately constant velocity. Virgo is the least massive of these three systems. Assuming that the CFs in NGC~7618 and UGC~12491 move at about the same velocity as in Virgo ($55 \Kpc/\Gyr$, see Fig.~7 in \citealt{Roediger2011}), the group-centric distance of 20 kpc of the CFs in both groups translates into an age of about 0.36 Gyr,  consistent with the first estimate. 

%***********************
\section{Substructure of cold fronts and implications for ICM properties}
%***********************
%
%FFFFFFFFFF
\begin{figure}
\centering
\includegraphics[trim=0 0 0 0,clip,angle=0,width=0.45\textwidth]{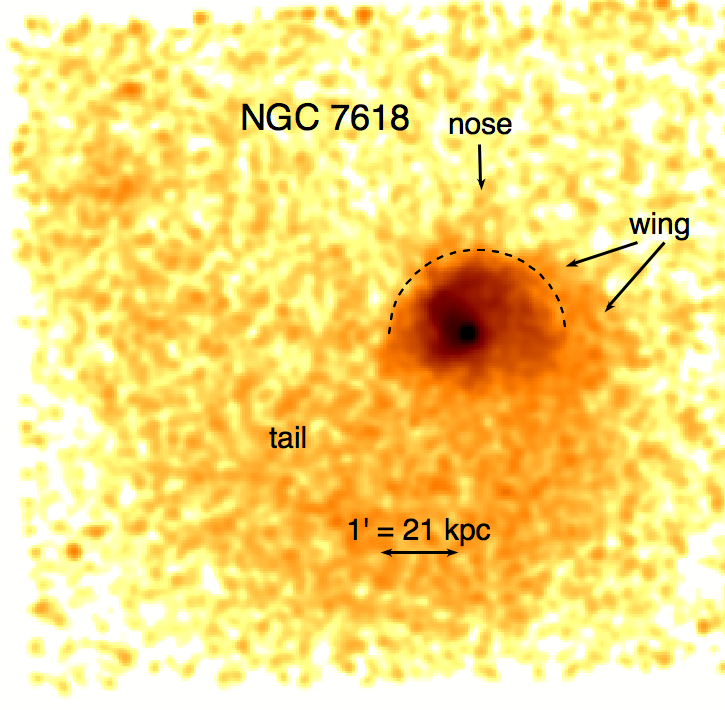}
\caption{Chandra/ACIS-S image of NGC~7618 in the 0.5-2.0~keV band, background-subtracted, exposure corrected, Gaussian-smoothed to 6 arcsec.  The logarithmic color scale is chosen to highlight the substructure of the cold front. Prominent features are labelled. The dashed arc marks the cold front.}
\label{fig:N7618}
\end{figure}
%FFFFFFFFFF
%
%FFFFFFFFFF
\begin{figure}
\centering
\includegraphics[trim=0 140 0 10,clip,angle=0,width=0.45\textwidth]{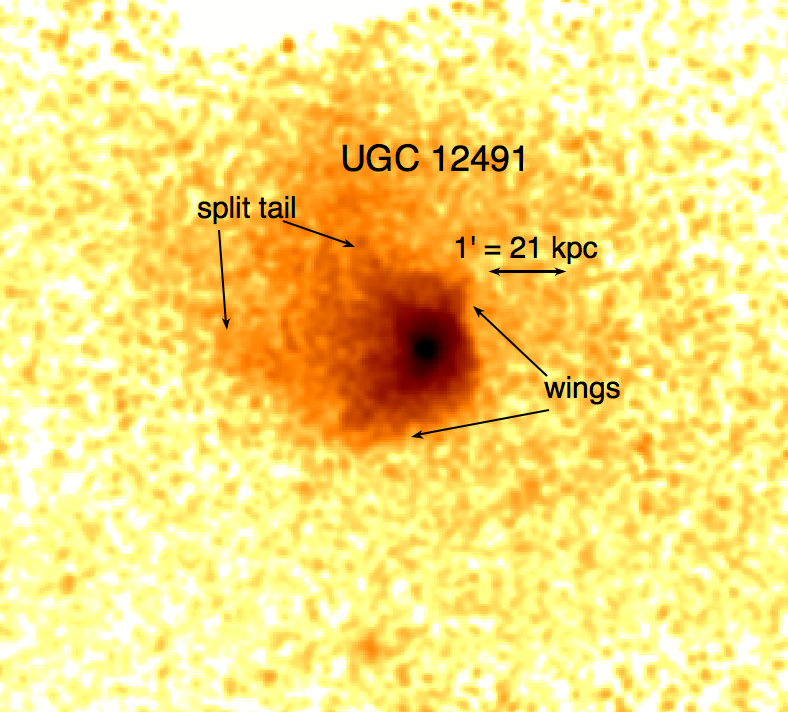}
\caption{Same as Fig.~\ref{fig:N7618} but for UGC~12491, Gaussian-smoothed to 4 arcsec.}
\label{fig:U12491}
\end{figure}
%FFFFFFFFFF
%
%FFFFFFFFFF
\begin{figure}
\centering
\includegraphics[trim=0 0 0 0,clip,angle=0,width=0.45\textwidth]{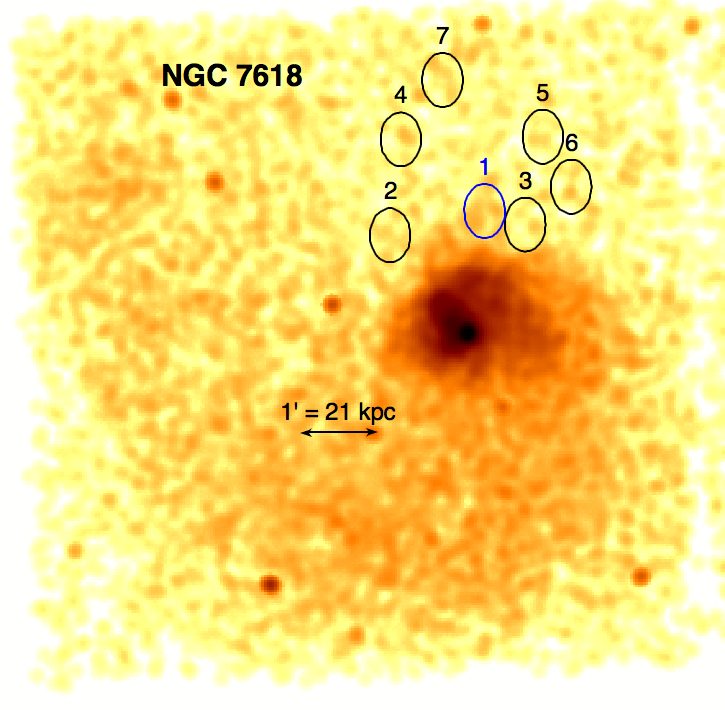}
\caption{Raw Chandra/ACIS-S image of NGC~7618 in the 0.7-1.4~keV band, Gaussian-smoothed to 6 arcsec.  The ellipse 1 covers the "nose" feature, ellipses 2 and 3 are at the same distance from the galaxy center, and ellipses 4 to 7 are placed such that they cover patches of  high background. All ellipses have the same size, the number of counts in them is listed in Table~\ref{tab:counts}.}
\label{fig:N7618_counts}
\end{figure}
%FFFFFFFFFF
%
%TTTTTTTTTTTTTT
\begin{table}
\caption{Number of counts in elliptical regions shown in Fig.~\ref{fig:N7618_counts}.}
\label{tab:counts}
%\begin{center}
\centering
\begin{tabular}{lccccccc}
\hline\hline
region & 1 (nose) & 2 & 3 & 4 & 5 & 6 &  7\\
\hline
counts & 60 & 42 & 34 & 34 & 29 & 33 & 37 \\
\hline
\end{tabular}
%\end{center}
\end{table}
%TTTTTTTTTTTTT
%
%
The most interesting features of NGC~7618 and UGC~12491 are the irregular shapes of their sloshing CFs. In Figures~\ref{fig:N7618} and~\ref{fig:U12491} we show the central regions for both galaxies where the color scale has been chosen to highlight the substructure along their CFs. The CF north of the nucleus of NGC~7618 displays a ``nose'' and a wing as labelled in Fig.~\ref{fig:N7618}. There may be more distortions/wings along the western edge of the tail, south of the structure labelled "wing". UGC~12491 has a smooth CF to the south-west which, however, terminates in two wings, one to the south-south-east, one to the north-north-west. Its tail appears to be split at about 30 kpc north-east from the nucleus and possibly again at 50 kpc to the north. Alternatively, these splits in the tail could be regarded as wings or distortions along the outside edge of the tail. In both groups, these CF substructures have linear scales of about 15 kpc. 

To demonstrate  significance of the nose feature at NGC~7618, we  compared the brightness of this feature with the adjacent background level.  We performed this analysis on the raw image in the 0.7 to 1.4 keV band, where the ratio of source to background counts has been optimized only by the choice of energy band.  We
 placed  the elliptical region 1 shown in Fig.~\ref{fig:N7618_counts} over the nose. Ellipses 2 and 3 are at the same distance from the galaxy center, and ellipses 4 to 7 are placed north of the CF. All ellipses have the same size. We ensured that all ellipses are not contaminated by the brighter emission inside the CF. The number of counts in each ellipse is listed in Table~\ref{tab:counts}. Ellipse 2 is the brightest of the background ellipses and contains 42 counts. Assuming Poisson noise implies a standard deviation of $\sigma=6.5$. The 60 counts in ellipse 1 on the nose feature are $2.7\;\sigma$ above the background level as defined by ellipse 2. The average background level of regions 2 to 7 is 35 counts, corresponding to $\sigma=5.9$. Therefore the counts in the nose regions are $4\;\sigma$ above this level. 
The length scale of the confidently detected structures is about 15 kpc. With the current data, smaller structures cannot be detected at comparable significance, because, assuming a comparable surface brightness also in smaller structures, the significance, i.e.~the excess of source counts above the background divided by the standard deviation of the background, is proportional to the considered length scale.
  To better define the CF shape, deeper observations are required.

We suggest that the distortions in these CFs are the result of KHIs, which are expected to arise due to shear flows along the CFs and are routinely seen in non-viscous hydrodynamical simulations (e.g.~Fig.~A1 in \citealt{Roediger2011}, also \citealt{Roediger2012a496,ZuHone2010}). As outlined in Sect.~\ref{sec:intro} above, both distorted CFs and smooth, arc-like ones have been observed, and the presence or absence of KHI-like distortions constrains the effective viscosity of the ICM and tangential magnetic fields along the fronts, which can both suppress the growth of the KHI. After a short remark on the effect of gravity we discuss these two ICM properties below.

Gravity suppresses the KHI (\citealt{Chandrasekhar1961}) at wavelengths above length scales of
%-----------
\begin{eqnarray}
 \lambda\Max &\approx & 18\Kpc \, \left(\frac{D}{1.5}\right)^{-1} \left( \frac{U}{200\Kms}\right)^{2}  \times \\
 &&  \left( \frac{g}{3\times 10^{-8}\cmss}\right)^{-1} \label{eq:tau0_value}\;\textrm{with} \; D =  \rho_1 / \rho_2,
\end{eqnarray}
%======
where we inserted typical values for the density contrast ($\rho_1$ and $\rho_2$ are the densities of the gas
interior and exterior to the CF, respectively) across the CF, $D$, and the shear velocity $U$, along the front. We  derived the gravitational acceleration $g$ from the azimuthally averaged ICM density and temperature profiles  in UGC~12491 given in the ACCEPT sample (\citealt{Cavagnolo2009}), assuming hydrostatic equilibrium. This length scale $\lambda\Max$ is comparable to the cluster-centric distance of the CFs, i.e.~the size of the low entropy core, which already limits the maximum perturbation length scale. Consequently, gravity does not play an important role in suppressing KHIs because the size of the cores is of the order of or smaller than $\lambda_{\mathrm{max}}$.

The KHI can be suppressed by sufficiently strong magnetic fields aligned with the interface.  The {\it presence} of KHI puts an upper limit on the magnetic field strength in these groups, because, if the combined magnetic pressures of the tangential magnetic fields at the hot and cold side of the CF, $(B\Cold^2 + B\Hot^2)/(8\pi)$ exceeds the limit (\citealt{Vikhlinin2002})
%--------------
\begin{eqnarray}
\frac{B\Cold^2 + B\Hot^2}{8\pi} 
&>&  0.14 \; p\ICM \left(\frac{\mathrm{M}}{0.5}\right)^2  \left(\frac{1+T\Cold/T\Hot}{1.5}\right)^{-1}
\end{eqnarray}
%========
the growth of the KHI should be suppressed ($p\ICM$ is the ICM pressure, $\mathrm{M}$ is the Mach number of the shear flow, and $T\Hot$ and $T\Cold$ are the temperatures at the warmer and colder side of the discontinuity, respectively). For an ICM pressure of  $p\ICM=10^{-3} \KeV\ccm$  at the CF this corresponds to a total magnetic field of  2 $\mu$G at the CF. The presence of fields just weaker than this limit increases the growth time of the instability, but only moderately. 
\citet{ZuHone2011} showed that gas sloshing typically amplifies the magnetic fields at the CFs by up to an order of magnitude, implying  initial field strengths of around $\sim$0.3 $\mu$G  in these two groups, which is within other observational limits for groups (e.g.~\citealt{Guidetti2010}).

\citet{Junk2010} consider the KHI in the presence of viscosity and derive the dispersion relation for a shear flow of relative velocity, $U$, between two layers of incompressible fluids with constant densities, $\rho_1$ and $\rho_2$ with $\rho_1 \ge \rho_2$.  They assume a constant kinematic viscosity, $\nu$, in both fluids. The corresponding growth timescale as a function of perturbation scale, $\lambda$, can be expressed as 
%-----------
\begin{eqnarray}
 \tau (\lambda) &=& \frac{1}{2}\tau_0 [\sqrt{1 + (\lambda / L)^2} + 1]\;\;\textrm{with}
 \label{eqn:tau_KH} \\
 L &=& \sqrt{\Delta}\;\frac{\pi \nu}{U}\;\;, \label{eq:KHlength}
 \tau_0 = \frac{\nu\Delta}{U^2} \;\;\textrm{and} \label{eq:tau0}\\
 \Delta &=& \frac{(\rho_1 + \rho_2)^2}{\rho_1 \rho_2} \approx \frac{\rho_1}{\rho_2} \;\;\textrm{for}\; \rho_1 \gg \rho_2. \nonumber
\end{eqnarray}
%======
The presence of viscosity separates two regimes of scale lengths above and below $L$. For $\lambda\gg L$ the viscosity is irrelevant and the instability grows at the same rate as in the inviscid case. For $\lambda < L$,  the growth timescale is approximately constant at $\tau\approx\tau_0$, whereas in the inviscid case the growth timescale scales linearly with $\lambda$. Thus, the viscosity slows down the growth rate compared to the inviscid case for $\lambda< L$. Using the definition of $L$ in Eqn.~\ref{eq:KHlength}, this inequality can be written as
%--------------
\begin{equation}
  \frac{\lambda U}{\nu}< \pi\sqrt{\Delta} ,
\end{equation}
%========
where $\frac{\lambda U}{\nu}$ is the Reynolds number associated with the perturbation length $\lambda$. Naturally,  the viscosity becomes relevant at small  Reynolds numbers.  In fact, the constant growth time for $\lambda< L$ derived above is true for the {\it onset} of the instability only. Numerical tests for the long-term evolution of the KHI in viscous gases show that the viscosity quickly erases the shear flow parallel to the interface and thus slows down and eventually prevents the growth of the instability at small Reynolds numbers (Roediger et al., in prep.). 
%
%TTTTTTTTTTTTTT
\begin{table*}
\begin{minipage}{\textwidth}
\caption{Full Spitzer viscosity should increase the growth time $\tau(\lambda)$ of the KHI for perturbations with length scales  $\lambda<L$, to $\tau=\tau_0$ (Eqn.~\ref{eqn:tau_KH}).  This table lists values for  the limiting  length scale $L$ and the resulting growth time $\tau_0$ at CFs in different clusters, along with assumed values for temperature $T$, density $n$ and shear velocity $U$ at each CF. Furthermore, we list for each CF the distance to the cluster/group center $r\CF$ and its age $\tau\CF$.}
\label{tab:examples}
%\begin{center}
\centering
\begin{tabular}{lccccccc}
\hline\hline
object & $T/$ & $n/ $ & $U/ $ & $L/$ & $\tau_0 /$ & $r\CF/$  & $\tau\CF/$ \\
 & keV & $10^{-3}\ccm$ & $\Kms$ & kpc & Myr & kpc  & Gyr \\
\hline
N7618/U12491 &  1.2 & 5 & 200 & 0.9 & 3 & 20 & 0.4\\
%U12491&&&&&&&\\
Virgo\footnote{\citealt{Roediger2011}} northern CF & 2.5 & 3 & 300 & 6 & 13 & 90  & 1.5 \\
A496\footnote{\citealt{Roediger2012a496}} northern CF & 4.2 & 8 & 400 & 6 & 10 & 60  & 0.7\\
A2142\footnote{\citealt{Markevitch2000}} south-east CF & 8 & 10 & 400 & 24 & 41  & 70 &  2\\
A2142 north-west CF & 8 & 2 & 600 & 81 & 91  & 360 & 2\\
\hline
\end{tabular}
\end{minipage}
%\end{center}
\end{table*}
%TTTTTTTTTTTTTT
%

We  calculate the numerical values for  $L$ and $\tau_0$ for CFs in different clusters in Table~\ref{tab:examples}. We  derive the kinematic viscosity $\nu=\frac{\mu}{\rho\ICM}$ from the Spitzer value for the  viscosity, $\mu$ (\citealt{Spitzer1956}), assuming a Coulomb logarithm of $\ln \Lambda=40$, ICM densities and temperatures as measured at each CF, and shear velocities either derived from simulations (for Virgo, A496) or assuming shear velocities of Mach 0.4, a typical value found in sloshing simulations. Additionally, we list for each CF its distance to the cluster center and its age as derived by simulations (for Virgo and A496) or from the cluster-centric distance and a constant CF velocity as described above. Abell 2142 is somewhat hotter than the massive cluster simulated by \citet{Roediger2012fastslosh}. Assuming that the CFs in Abell 2142 move outwards with the same velocity as in this simulated cluster, we estimate that the sloshing in Abell 2142 was triggered about 2 Gyr ago. Sloshing CFs generally span more than 90 degrees in position angle and thus more than their $r\CF$ in azimuthal extent.

The estimates of $L$ and $\tau_0$ (see Table~\ref{tab:examples}) imply that viscosity does not significantly suppress the growth of KHIs in groups and poor clusters, because it can affect only perturbations at spatial scales significantly smaller than the CF radius or azimuthal extent. Also the estimated growth times are much shorter than the age of the CFs. Consequently, we should observe KHIs in NGC~7618, UGC~12491 and A496, which all show distorted fronts. 
In NGC~7618 and UGC~12491, the KHIs may grow even more rapidly, because in this roughly equal mass merger the shear flows along the fronts may be stronger than the  Mach $\sim$0.4 assumed here, which are typical for minor mergers.
The northern front in Virgo should also exhibit KHIs, which may be hidden in the low spatial resolution of the current shallow XMM-Newton observation (\citealt{Simionescu2010}). The estimated values of $L$ and $\tau_0$ for the hot cluster Abell 2142 are largest both in absolute terms and in comparison to the CF radii, and the viscosity is expected to have the strongest effect here. Indeed, there is no obvious evidence for KHI in the current Chandra data.

A similarly diverse picture arises regarding the structure or smoothness of merger CFs, i.e.~the interfaces between the atmospheres of a cluster and an infalling subcluster or galaxy.  \citet{Vikhlinin2002} find that a tangential magnetic field of about 10 $\mu$G is required to prevent KHI at the smooth merger CF in Abell 3667. \citet{Dursi2008} have shown that such magnetic draping layers can form. However, \citet{Churazov2004} argue that the absence of the KHI near the stagnation point can be due to a small, unobservable intrinsic width of the curved front and the spatial variation of the shear flow along it. \citet{Mazzotta2002} suggest that the wings of the merger CF in A3667 indeed break up in KHIs. Recent deep Chandra and XMM observations of the merger fronts at the elliptical galaxies NGC 4552 (\citealt{Machacek2006a}) and NGC 4472 (\citealt{Kraft2011}) falling into the Virgo cluster reveal significant substructure, i.e.~horns, kinks and noses for both. In contrast, the  upstream edge of NGC 1404 (\citealt{Machacek2005}) falling into the Fornax cluster is not distorted. Like the sloshing CFs, some merger CFs are also distorted while others are not. The reason for this difference is yet unclear, it may be of dynamical nature, e.g.~the amplitude of shear flow velocity or density contrast, or depend on the magnetic fields or transport coefficients of the ICM.

Our simple estimate regarding the growth timescales of KHIs in the linear regime neglects several important issues that are relevant for real CFs: (i) The interface is not plane-parallel but curved, and the shear flow along it may vary with position and time. This is especially true for the merger CFs. In sloshing CFs, the interface moves away from the cluster center, stretching growing perturbations. This increases the growth time because the perturbation length increases and because the density at the interface decreases. (ii) The growth time scales refer to the initial linear growth phase only. The evolution of CFs takes place on much longer timescales, where the KHI  enters the non-linear regime. (iii) For a Spitzer-like viscosity, the kinematic viscosity is not constant across the interface, but the ratio of Reynolds numbers, between the hot and cold layer, scales with temperature contrast to the power of 3.5. (iv) The effect of tangential magnetic fields and viscosity may combine in suppressing the KHI.

The recent and upcoming deep and high resolution observations of merging and sloshing cold fronts provide us with a wealth of information regarding the presence or absence of KHIs at both sloshing and merger CFs and hence the amplitude of magnetic fields and viscosity in the ICM. Understanding the data requires a systematic study of group and cluster CFs to determine which conditions lead to the growth or suppression of the KHI. The list of complications given above makes clear that simple 1 or 2D analogues will not be sufficient to precisely interpret the observations. Additionally, both the required magnetic field strengths and viscosity for the suppression of the KHI depend also on the dynamical conditions like local density and shear velocity.  Simulations  tailored specifically to the object in question are required to correctly interpret the data. The direct comparison of the overall properties of the CFs, i.e.~positions and contrasts across them, allows the merger history and thus the current and recent dynamical state of the ICM to be constrained, e.g.~densities, temperatures and shear flow histories at the CFs. Consequently, these dynamical properties can be disentangled from the ICM viscosity and magnetic field strengths, and these can be measured. We will follow this strategy in a series of forthcoming papers.

%**********************
\section*{Acknowledgments}
%**********************
E.R.~acknowledges  support by the Priority Programmes 1177 ("Witnesses of Cosmic History") and 1573 ("Physics of the Interstellar Medium") of the DFG (German Research Foundation),  the supercomputing grants NIC 3711 and 4368 at the J\"ulich Supercomputing Center, a visiting scientist fellowship of the Smithsonian Astrophysical Observatory, and the hospitality of the Center for Astrophysics in Cambridge. 
M.E.M.~acknowledges support by the Nasa grant NNX07AH65G. 
This work was supported by NASA grant NAS8-03060.
We thank the referee for her/his helpful and constructive comments.

%*******************************************************************
%*************** R E F E R E N C E S *******************************
%*******************************************************************
%

\end{document}